\documentclass[%
 reprint,
 amsmath,amssymb,
 aps,
]{revtex4-2}

\usepackage{graphicx,amsmath}
\usepackage{dcolumn}
\usepackage{bm}
\usepackage[skins,theorems]{tcolorbox}
\usepackage{amssymb}
\usepackage{tablefootnote}
\usepackage{units}
\usepackage{threeparttable,lipsum}
\usepackage{mathtools}
\usepackage{hyperref}
\usepackage{dcolumn}
\usepackage{bm}
\usepackage{aas_macros}

\tcbset{highlight math style={enhanced,
 colframe=cyan!10!gray,colback=gray!5!white,arc=4pt,boxrule=1pt,
  drop fuzzy shadow}}

  {\empheq[box=\tcbhighmath]{align}}
  {\endempheq}

\begin{document}

\title{White Paper on the TUNL Nuclear Astrophysics Program}

\author{Christian Iliadis}
\author{Art E. Champagne}
\author{Akaa D. Ayangeakaa}
\author{Robert V. F. Janssens}
\affiliation{Triangle Universities Nuclear Laboratory (TUNL), Duke University, Durham, North Carolina 27708, USA}
\affiliation{Department of Physics \& Astronomy, University of North Carolina at Chapel Hill, NC 27599-3255, USA}

\author{Richard Longland}
\affiliation{Triangle Universities Nuclear Laboratory (TUNL), Duke University, Durham, North Carolina 27708, USA}
\affiliation{Department of Physics, North Carolina State University, Raleigh, NC 27695, USA}

\begin{abstract}
The White Paper describes the nuclear astrophysics program at the Triangle Universities Nuclear Laboratory (TUNL), with the intent of providing input for the 2023 NSAC Long Range planning process.
TUNL is operated jointly by North Carolina Central University, North Carolina State University, The University of North Carolina at Chapel Hill, and Duke University. TUNL houses three world-class facilities for nuclear astrophysics research: the Laboratory for Experimental Nuclear Astrophysics (LENA); the Enge Magnetic Spectrograph; and the High-Intensity $\gamma$-ray Source (HI$\gamma$S). We discuss past successes, the present status, and future plans.
\end{abstract}

\date{\today}

\maketitle


\section{\label{sec:intro}Program Overview}
In a 2013 comparative review, the nuclear astrophysics program at the Triangle Universities Nuclear Laboratory (TUNL) was ranked in the top two among 32 university and national laboratory groups in the nation supported by the U.S. DOE. Over this time period, about ten students received their Ph.D.'s for experimental work in nuclear astrophysics at TUNL. The program has a holistic foundation, addressing diverse themes in nuclear astrophysics and forging their interconnections. This is illustrated in the coupling of laboratory measurements, which will be discussed below, with our leadership in developing numerical tools for cutting edge research. During the past decade, examples of ``firsts'' include: the estimation of Monte Carlo thermonuclear reaction rates \cite{Longland:2010is}, Bayesian estimates of astrophysical $S$ factors \cite{iliadis16}, the construction of a next-generation library of thermonuclear reaction rates (STARLIB) \cite{Sallaska2013}, Monte Carlo reaction network calculations \cite{Longland2012,Iliadis:2015gp}, Genetic Algorithms applied to Big Bang nucleosynthesis \cite{IC20}, Bayesian methods for analyzing particle transfer measurements \cite{Marshall2019,Marshall2020}, and others. All of the developed tools are freely available and are now frequently used by the community. One recent example is the adoption of STARLIB by the Max-Planck-Institute group to study black-hole mergers and gravitational-wave emission \cite{Farmer_2019}. 

TUNL is a place that generates many new ideas, and provides motivation for new measurements to the community, involving both stable and radioactive ion beams. By performing large-scale nucleosynthesis simulations, key nuclear reactions are being identified which crucially impact our understanding of classical nova explosions \cite{downen2013,Kelly_2013}, globular star clusters and early galaxy evolution \cite{Iliadis_2016}, the origin of presolar stardust grains in meteorites \cite{iliadis2018}, and others. These simulations provide a solid foundation for laboratory measurements world-wide.  

As a highly ranked program in the nuclear structure and astrophysics portfolio of the U.S. DOE, TUNL recently secured funding (about $\$3$M) to upgrade the Laboratory for Experimental Nuclear Astrophysics (LENA). Future plans for LENA are discussed in Sec.~\ref{sec:lena}. For many nuclear reactions, thermonuclear fusion occurs at kinetic energies too small for direct access in the laboratory. Such reactions are being measured indirectly using transfer reactions at the TUNL Enge Spectrograph. Very few other laboratories can match the Enge's superb energy resolution and large solid angle, and, consequently, this facility is complementary to LENA. Recently, TUNL was also awarded about $\$1.5$M to upgrade the low-energy injector system at the Tandem laboratory to fully capitalize on these capabilities. The Enge Spectrograph is further discussed in Sec.~\ref{sec:enge}. Yet another complementary and unique facility at TUNL is the High Intensity Gamma-Ray Source (HI$\gamma$S). It allows for the excitation of astrophysically important levels using linearly-polarized and quasi-monoenergetic $\gamma$-ray beams. The decay of the excited levels is highly anisotropic \cite{iliadis21} and the measured angular correlations provide a wealth of nuclear structure information on astrophysically important low-spin states. The HI$\gamma$S program is discussed in Sec.~\ref{sec:higs}.

During the past decade, the TUNL nuclear astrophysics group published about $73$ peer-reviewed papers in top journals, with $9500$ total citations. TUNL researchers have also published the only nuclear astrophysics textbook in print \cite{Iliadis:2015ta}, which is used to train advanced undergraduate and graduate students world-wide.
\begin{figure*}[t]
\includegraphics[width=1.0\textwidth]{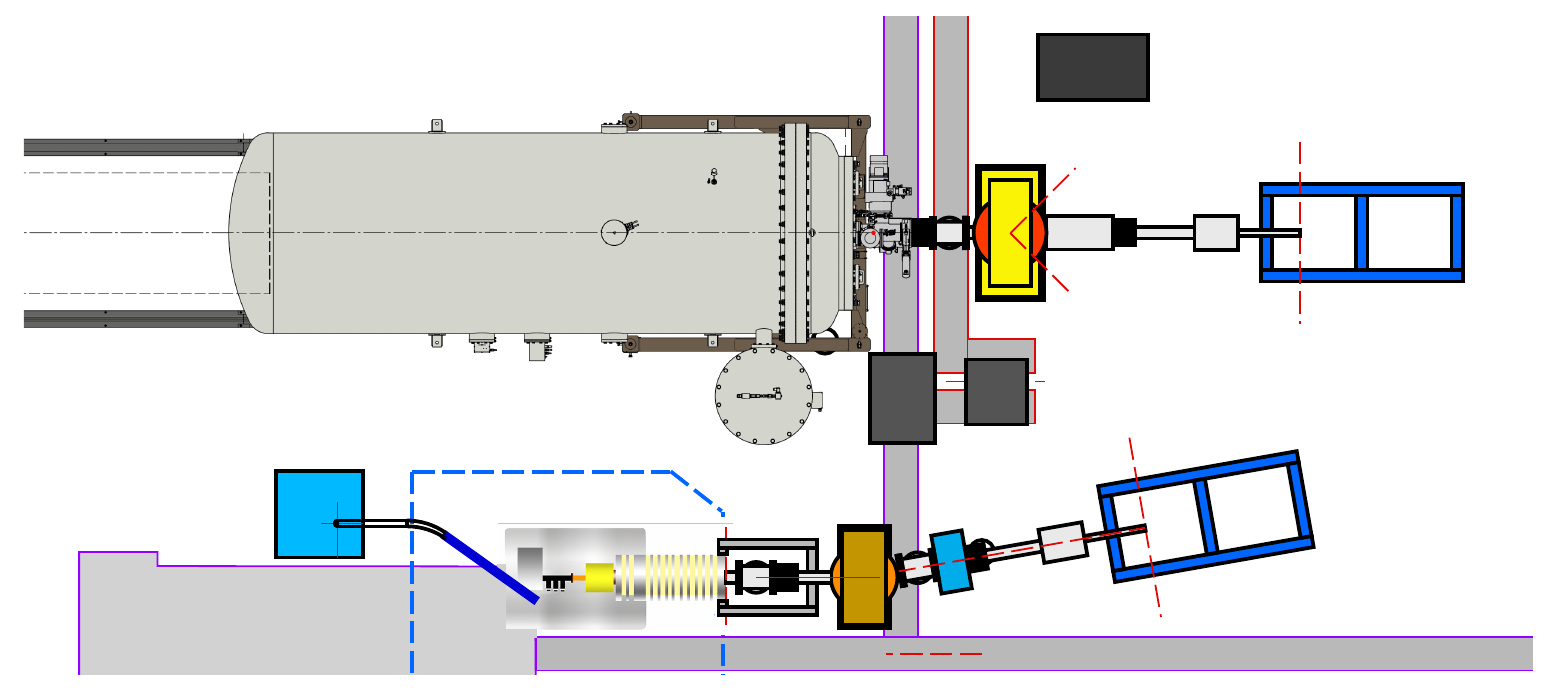}
\caption{Floor plan of the LENA II facility. The $2$-MV Singletron and the ECR accelerator are depicted at the top and bottom, respectively, with each machine injecting ions into a dedicated beam line. This new facility will deliver beams of exceptionally high intensity and time structure for experiments dedicated to nuclear astrophysics.}
\label{fig:lena2}
\end{figure*}
%

\section{\label{sec:lena}The Laboratory for Experimental Nuclear Astrophysics (LENA)}
The Laboratory for Experimental Nuclear Astrophysics (LENA) is a dual-accelerator facility designed to measure directly nuclear reactions of astrophysical interest, preferably at the energies of stellar burning. It is undergoing a major upgrade in 2022. We will refer to the status of the laboratory before and after this year as LENA~I and LENA~II, respectively.

LENA~I housed two accelerators: a 230-kV ECR accelerator and a 1~MV JN Van de Graaff. A unique feature of the former has been its high current, which, when combined with a $\gamma$-ray coincidence spectrometer, enables measurements at sensitivities that rival what can be achieved at existing underground facilities. A prominent example is the measurement of the $^{17}$O(p,$\gamma$)$^{18}$F reaction \cite{buckner2015}, which is of crucial importance for the interpretation of observed oxygen elemental and isotopic abundances in stellar spectra. Both primary and secondary $\gamma$-ray transitions from this reaction were measured at LENA~I, down to a center-of-mass energy of $160$~keV, with an unprecedented proton intensity of $2$~mA. This reaction was also measured at the LUNA underground facility at Gran Sasso \cite{PhysRevC.89.015803}, where both primary and secondary $\gamma$-ray transitions were detected down to $200$~keV. The proton beam intensity at LENA~I exceeded that of LUNA by about an order of magnitude, allowing for a cross section measurement at lower energies. The TUNL work was featured as a ``Science Highlight'' on the website of the U.S. DOE.

Other highlights from LENA~I are the measurements of $^{29}$Si(p,$\gamma$)$^{30}$P \cite{Downen_2022,PhysRevC.105.055804}, important for interpreting presolar grains from classical nova explosions, and $^{30}$Si(p,$\gamma$)$^{31}$P \cite{dermigny20}, important for explaining abundance correlations observed in globular cluster stars. For both reactions, the new TUNL data improved the thermonuclear reaction rates significantly.

The new facility, LENA II (see Fig.~\ref{fig:lena2}) will also feature two accelerators. The ECR accelerator has been upgraded with a new pulsing system, a solenoid magnet to provide the ECR field, and a novel field clamp in the extraction region to improve beam brightness. The goal of this upgrade is to produce H$^{+}$ beam currents of 20~mA with the capability of pulsing at about 10\% duty cycle. Pulsing will reduce cosmic and environmental backgrounds by a factor of $10$ without reducing beam intensity relative to what was achieved with LENA I. The ECR accelerator will be paired with a novel 2-MV Singletron accelerator, which is designed and built by High Voltage Engineering Europa B.V. The Singletron has been built specifically for LENA and represents an entirely new concept. It pairs an ECR ion source with a fast buncher/chopper system in the terminal. It has been shown to produce high beam currents (DC currents for H$^{+}$ and He$^{+}$ of approximately 0.5 $-$ 2 mA for $E$ $<$ $1$~MeV and 2 mA for $E$ $>$ $1$~MeV), with pulsing frequencies up to 4 MHz and pulse widths of 2 – 20 ns. Pulsing can again be used for background reduction, as well as for neutron spectroscopy via time-of-flight techniques. {\it Measurements at LENA II will investigate, in the near future, helium burning reactions and neutron sources for the astrophysical $s$ process.}

\section{\label{sec:enge}The Enge Magnetic Spectrograph}
\begin{figure}
\centering
\includegraphics[width=0.45\textwidth]{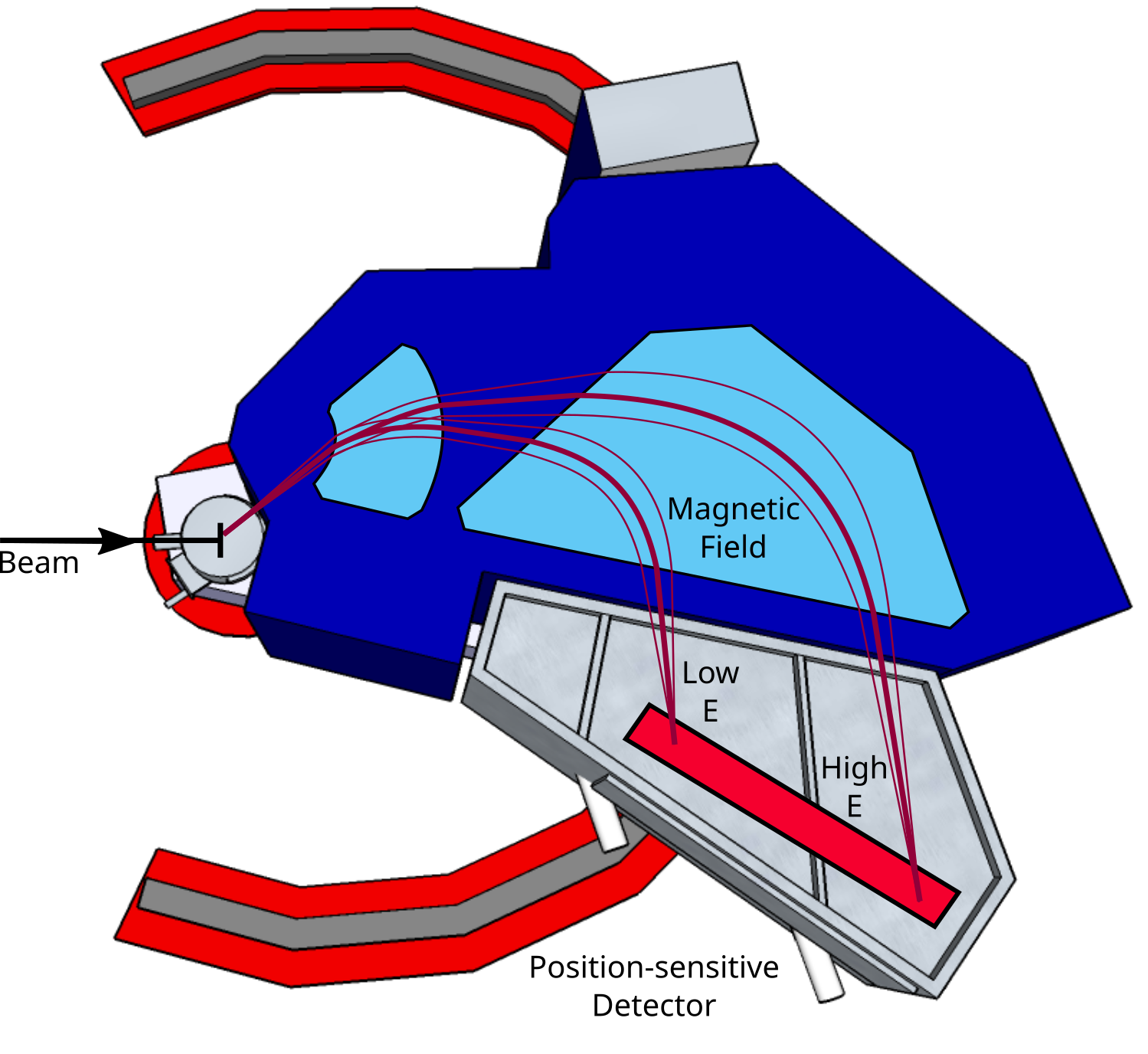}
\caption{Particle trajectories through TUNL's high-resolution Enge split-pole spectrograph. The instrument has a superb energy resolution and solid-angle acceptance, enabling the study of astrophysically important nuclear levels that cannot be measured directly.}
\label{fig:EngeTrajectories}
\end{figure}

Many resonances cannot be measured directly in charged-particle reactions, especially when they are located close to the proton or $\alpha$-particle thresholds. Their reaction rate contributions must be estimated from nuclear structure properties obtained in transfer reaction studies (i.e., excitation energies, spins, parities, spectroscopic factors, asymptotic normalization coefficients, and branching ratios). The high-resolution Enge split-pole spectrograph at TUNL \cite{Spencer1967} has been designed for providing such data. The spectrograph is located in the Tandem Laboratory. Particle beams are delivered to the spectrograph scattering chamber through a high-resolution beam line, where two back-to-back 90$^{\circ}$-analyzing magnets provide accelerator feedback control \cite{Wilkerson1987}. Particles emitted from the reaction enter the split-pole magnet, are momentum-analyzed, and reach the focal plane, where their position is related to magnetic rigidity, and, hence, resonance energy (Fig.~\ref{fig:EngeTrajectories}). The measured angular distributions are used to determine the spin-parity and spectroscopic factor of important excited states. 

A number of novel features make the TUNL split-pole spectrograph a world-class facility for nuclear astrophysics. Measuring the properties of important states requires a high resolving power from both neighboring levels and contaminant states. The combination of a high beam energy resolution and a small beam spot of 1~mm on target results in a theoretical energy resolution at the focal plane dominated only by target straggling effects (i.e., a few tens of keV). The Enge focal-plane detector has been designed to achieve an exquisite position resolution of approximately 0.3~mm \cite{Marshall2019}. Two position-sensitive sections facilitate digital reconstruction of the kinematic focal plane. Maximum particle resolution can be achieved at all times, even when mechanical positioning of the focal-plane detector is not optimized. Multiple position sections also enable the user to perform detector-angle sub-sampling, providing outstanding performance for very high-resolution scattering angle measurements. Developments are currently underway to add a silicon detector array for measuring particle decay branching ratios. Upgrades of the Tandem's ion sources are in progress and will provide a factor of 10 increase in beam intensity, significantly improve beam stability, and unlock new capabilities using heavier ion beam species. These upgrades are further cementing the TUNL split-pole spectrograph facility as a world leader in particle transfer measurements for nuclear astrophysics.

The first experimental results from the new TUNL spectrograph facility made headway in understanding the abundances in the Ar-Ca region observed in classical nova explosions. By measuring the $^{40}$Ca($^3$He,$\alpha$)$^{39}$Ca pickup reaction \cite{Setoodehnia2018}, the rate uncertainties of the astrophysically important $^{38}$K(p,$\gamma$)$^{39}$Ca reaction were significantly reduced. This was followed by another study of excited states in $^{35}$Cl important for the $^{34}$S(p,$\gamma$)$^{35}$Cl reaction rate~\cite{Setoodehnia2019}. The $^{23}$Na($^3$He,d)$^{24}$Mg proton transfer reaction was used to determine the energies, spin-parities, and spectroscopic factors of states in the astrophysically important $^{23}$Na(p,$\gamma$)$^{24}$Mg reaction. This latter is key to understanding the puzzling sodium-oxygen anti-correlation observed in globular clusters \cite{Cassisi2020}. New Bayesian statistical methods were developed to modernize the focal-plane energy calibration \cite{Marshall2019}, leading to a factor of $2$ increase in the reaction rate~\cite{Marshall2021}. This work was featured as a highlight on the U.S. Department of Energy's website~\cite{MarshallDOEHighlight}. The Bayesian methods developed at TUNL for the data analysis using Distorted Wave Born Approximation (DWBA) theory \cite{Marshall2020} are available to the broader community\footnote{See: https://github.com/dubiousbreakfast/pfunk}. 

The TUNL Enge split-pole spectrograph is one of very few magnetic spectrographs worldwide that are dedicated to nuclear astrophysics measurements. For high-resolution studies, it has emerged as a leading device. A number of international collaborations have resulted from this program, e.g., with groups from the United Kingdom performing high-resolution particle spectroscopy measurements at TUNL \cite{Hamill2020,Frost-Schenk2022}. The latter of these two references achieved an energy resolution of less than 10 keV, and spurred additional measurements using the HELIOS spectrometer at ATLAS. This cross-pollination of nuclear astrophysics research efforts between TUNL and national user facilities is one example of the impact of university-based programs on the nuclear astrophysics community as a whole. {\it In the near future, this facility will focus on measurements important for classical novae explosions.}
%

\section{\label{sec:higs}The High-Intensity Gamma-Ray Source (HI$\gamma$S)}
Two world-class accelerator-based photon sources are located at TUNL: a storage ring Free-Electron Laser (FEL) and the High Intensity Gamma-ray Source (HI$\gamma$S). The former is a unique storage ring oscillator FEL, with the world’s longest laser cavity (about 54 meters) with an active medium that produces coherent photon beams in a wide range of wavelengths from IR (1060 nm) to VUV (190 nm). The latter is the highest-flux Compton $\gamma$-ray source in the world. It generates linearly or circularly polarized $\gamma$-ray beams with energies ranging from 1 to 120 MeV, with typical beam intensities of $\approx$ 10$^9$ s$^{-1}$ at $5$~MeV $\gamma$-ray energy. The energy resolution is typically 3\% in this energy range. The unique combination of high $\gamma$-ray flux, quasi-monoenergetic beam energy, and nearly 100\% beam polarization allows for nuclear astrophysics measurements that cannot be performed anywhere else.

Three broad types of nuclear astrophysics experiments can be performed at HI$\gamma$S: direct measurements that take advantage of reciprocity, nuclear resonance fluorescence (NRF) studies, and measurements of the photon strength function that improve predictions of statistical (i.e., Hauser-Feshbach) models. These will be described briefly below.
\begin{figure}[tp!]
\centering
\includegraphics[width=0.48\textwidth]{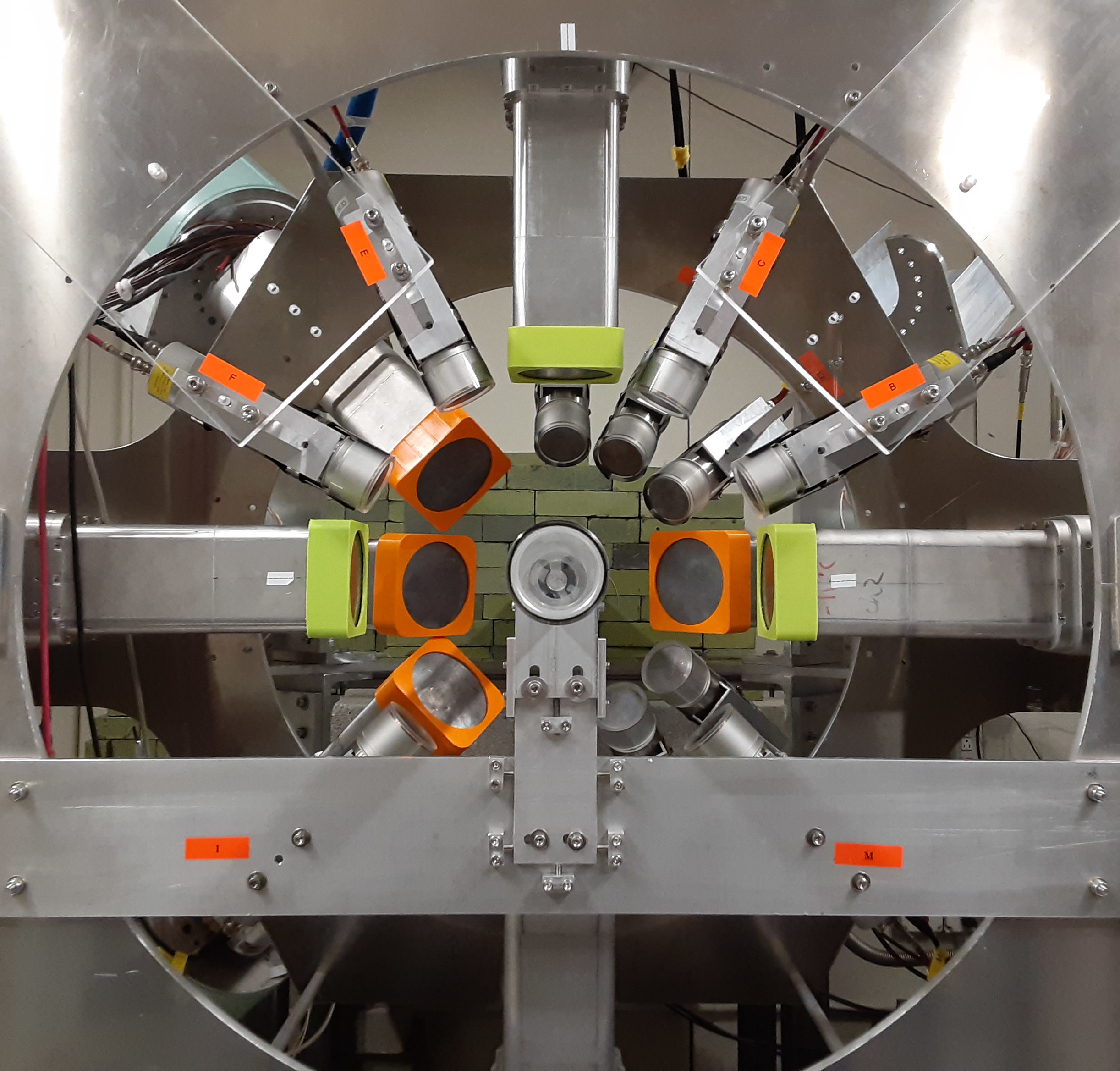}
\caption{The Clover Array at HI$\gamma$S, consisting of 8 clover-type HPGe detectors and 12 CeBr scintillation counters. The detectors are placed around the sample for in-beam NRF studies. The linear polarization of the incident $\gamma$-ray beam gives rise to distinct angular correlations that are highly selective for determining spins and parities of excited nuclear levels \cite{iliadis21}. The beam moves from the center of the figure towards the reader.}
\label{fig:clover}
\end{figure}

Neutrino-driven wind models of the $r$ process \cite{Iliadis:2015ta} predict that the $\alpha$ $+$ $\alpha$ $+$ $n$ $\rightarrow$ $^9$Be reaction represents the first step in the reaction chain that produces the seed nuclei \cite{Iliadis:2015ta}. Since this reaction cannot be measured in the laboratory, instead, the reverse process, $^9$Be($\gamma$,n)2$\alpha$, was studied at HI$\gamma$S \cite{PhysRevC.85.044605} and the rate of the forward reaction was obtained using the reciprocity theorem. This experiment achieved a $\gamma$-ray beam energy resolution of only $19$~keV in the threshold region, allowing for a much improved cross section measurement. Based on these results, it was found that the total reaction rate increased by $\approx$40\%, which impacts significantly predictions of the $r$ process. 

Half of the elements beyond iron is made by slow neutron capture nucleosynthesis ($s$ process), where the $^{22}$Ne($\alpha$,n)$^{26}$Mg reaction represents one of the two most important neutron sources \cite{Iliadis:2015ta}. A $^{26}$Mg level near the $\alpha$-particle threshold ($E_r^{c.m.}$ $\approx$ $630$~keV), presumed to have $J^{\pi}$ $=$ $1^-$ quantum numbers, was thought to dominate the rate in the astrophysically important region. This level was studied at HI$\gamma$S via NRF using a linearly polarized beam \cite{PhysRevC.80.055803}. The resulting angular correlation \cite{iliadis21} unambiguously determined the spin-parity as 1$^+$ (unnatural parity), which reduced the reaction rate by orders of magnitude. A similar study was recently performed at HI$\gamma$S \cite{PhysRevC.106.014308} to better understand the puzzling abundance correlations in the globular cluster NGC 2419. A $\gamma$-ray detection setup (Clover Array) for NRF studies at HI$\gamma$S is depicted in Fig.~\ref{fig:clover}.

In a number of astrophysical scenarios, thermal excitations of the target or residual nucleus dramatically impact the reaction rates. This occurs, for example, in the $s$ process, particularly when branch point nuclides are encountered on the nucleosynthesis path, and during the $p$ process that is driven by ($\gamma$,n) reactions \cite{Iliadis:2015ta}. In such cases, the reaction rates must be estimated using the nuclear statistical (Hauser-Feshbach) model, which requires as input, apart from optical model parameters and level densities, the $\gamma$-ray strength functions. The latter can be measured at HI$\gamma$S, taking advantage of the superb energy resolution and polarization of the incident $\gamma$-ray beam. Such studies have led, for example, to improved predictions of the $s$-process branching at $^{85}$Kr  \cite{PhysRevLett.111.112501}.

The unique experimental capabilities of HI$\gamma$S are also of significant interest to outside user groups. A prominent example is the measurement of the $^{16}$O($\gamma$,$\alpha$)$^{12}$C reaction \cite{osti_1825137} to constrain the rate of the inverse $^{12}$C($\alpha$,$\gamma$)$^{16}$O reaction, which is key for constructing improved models of stellar evolution and explosions. Another example is the model-dependent determination of the dipole response of $^{66}$Zn \cite{Savran2022}, which demonstrates the superiority of measurements with HI$\gamma$S beams over others using bremsstrahlung. {\it In the near-term future, experiments at HI$\gamma$S will focus on structure studies crucial for globular cluster nucleosynthesis.}

\begin{acknowledgments}
This work was supported in part by the DOE, Office of Science, Office of Nuclear Physics, under grants DE-SC0017799, DE-SC0023010, DE-FG02-97ER41041 (UNC), DE-FG02-97ER41042 (NCSU), and DE-FG02-97ER41033 (TUNL).
\end{acknowledgments}

\bibliography{paper}

\end{document}